\documentstyle[aps,prc,eqsecnum,epsfig,preprint,tighten]{revtex}

\begin{document}

   \newcommand{\ic}{\;\;\; ,}
   \newcommand{\ip}{\;\;\; .}
   \newcommand{\sfrac}[2]{\textstyle \frac{#1}{#2} \displaystyle}

\draft

\title{A Poincar\'e-Covariant Parton Cascade Model for
       Ultrarelativistic Heavy-Ion Reactions
       }
\author{V.\ B\"orchers
                     \thanks{Now at kidata AG,
                             K\"onigswinter, Germany},
        J.\ Meyer    \thanks{Now at Institut f\"ur Theoretische Physik,
                             Universit\"at Heidelberg, Germany},
        St.\ Gieseke \thanks{Now at II.Institut f\"ur Theoretische
                             Physik, Universit\"at Hamburg, Germany},
        G.\ Martens  \thanks{Now at Institut f\"ur Theoretische Physik,
                             Universit\"at Giessen, Germany},
        C.C.\ Noack
       }
\address{Institut f\"ur Theoretische Physik \\
         Universit\"at Bremen, D--28334 Bremen \\[3ex]
        }
\date{submitted to Phys.\ Rev.\ C}
\maketitle

\begin{abstract}
We present a new cascade-type microscopic simulation of nucleus-nucleus
collisions at RHIC energies. The basic elements are partons (quarks and
gluons) moving in $8N$-dimensional phase space according to
Poincar\'e--covariant dynamics. The parton-parton scattering cross
sections used in the model are computed within perturbative QCD in the
tree-level approximation. The $Q^2$ dependence of the structure
functions is included by an implementation of the DGLAP mechanism
suitable for a cascade, so that the number of partons is not static, but
varies in space and time as the collision of two nuclei evolves. The
resulting parton distributions are presented, and meaningful comparisons
with experimental data are discussed.
\end{abstract}

\pacs{03.30.+p,24.10.Lx,24.85.+p,25.75.-q}

\section{Introduction}                 \label{sec:intro}

Recent years have shown an increased interest in the study of heavy ion
reactions with projectiles and targets ranging all the way up to
Uranium, and laboratory energies up to 200 A$\cdot$GeV
\cite{QM96,QM97,QM99}. The RHIC collider at Brookhaven National
Laboratory, dedicated to ultrarelativistic heavy ion reactions, became
operational this year, and this heralds yet another new and exiting
stage of experiments, with the prospect of finally confirming the
signatures of a phase transition to the quark gluon plasma found at CERN
(cf.\ \cite{Hei00}).

In a theoretical microscopic description of such reactions it is
imperative to take into account the quark and gluon degrees of freedom,
even when one does not assume a phase transition to occur, and various
such microscopic models (generically called ``parton cascades'') have
been studied \cite{Wan91,Gei92,Wer93,Gei95,Sor95,Li95,Win96,Gei97}. In
as far as these models describe the motion of individual particles in
phase space, they are necessarily \emph{classical} models and thus
suffer in various degrees from the consequences of the
No-Interaction-Theorem \cite{Cur63},  which severely restricts the
possibility of ensuring the full Poincar{\'e} covariance in such models.
Indeed the models just mentioned exhibit their non-covariance by an
explicit dependence on the coordinate system in which the simulated
reactions are run.

In contrast, we present a parton cascade model which is formally
strictly Poincar{\'e}-covariant. This is not to say that our model is
free from the basic problem inherent in any parton description to date:
the very definition of the incoming nucleons in terms of their parton
content depends on the momentum scale used and thus seems to depend
unavoidably on the observer frame of reference in which the paricipant
nucleons are seen. We shall address this aspect of our model in detail
below (cf.\ Sect.\ \ref{sec:evolution}).

The paper is organized as follows. In Sect.\ \ref{sec:dynamic} we present
the details of our covariant formalism together with a description of
the basic cascade algorithm used. Sect.\ \ref{sec:ini_state} deals with
the construction of the initial state of the model, and
Sect.\ \ref{sec:parton_scattering} describes the partonic scattering
processes during the nuclear reaction. In Sect.\ \ref{sec:evolution} we
discuss the question of `parton evolution' as implemented in our code.
In Sect.\ \ref{sec:results} we present some numerical results and compare
them to experimental data if available. Sect.\ \ref{sec:conclusions}
contains a discussion and our conclusions.
\section{The dynamics of our model}    \label{sec:dynamic}

The No-Interaction-Theorem by Currie et al.\ asserts that the only
canonical Hamiltonian theory of $N$ particles which is
Poincar\'e-covariant is one in which all particles are free
\cite{Cur63}. One way to circumvent the consequences of this theorem is
to formulate the theory in 8$N$-dimensional phase space, i.e.\ in terms
of 4-vectors for the positions of the particles as well as for their
momenta:
\[ x_i := (t_i,\vec{r}_i), \text{\qquad} p_i := (E_i,\vec{p}_i),
          \text{\qquad} i = 1,\dots N \ip \]
These 4-vectors are taken to be functions of a
\emph{Poincar\'e-invariant} dynamical evolution parameter $s$, the
motion of the particles being determined by the set of Hamilton's
equations
\begin{eqnarray*}
 \frac{d}{ds}x_i(s) &=& \{H,x_i\} = -\frac{\partial H}{\partial p_i}  \\
 \frac{d}{ds}p_i(s) &=& \{H,p_i\} = +\frac{\partial H}{\partial x_i} \ic
\end{eqnarray*}
where the Hamiltonian $H$ as well as the interaction ``quasipotential''
$V$ are \emph{Poincar\'e-invariants}:
\[ H = \sum_{i=1}^{N} \frac{m_i^2 - p_i^2}{2m_i} \;
      +V(r_1, \dots, r_N; p_1 \dots ,p_N) \ip         \]

The details of such a dynamical theory have been described elsewhere
\cite{Pet94}. Here we emphasize two important features: (a) the
parameter $s$ governs the dynamical evolution of the system, but has no
further direct physical interpretation, (b) particles are (classically!)
off-shell -- $(p_i)^2 \neq (m_i)^2$ -- whenever they are within the
range of the quasipotential.

With an appropriately chosen attractive force (representing a
string-like interaction), this framework allows a description of hadrons
as bound systems of classical particles (``partons'') \cite{Beh94}.

In setting up a parton cascade, we use a drastically simplified
Hamiltonian which is in the spirit of previous hadronic cascade models
\cite{Cug82,Kit84,Pan92}: we take the interactions of the model to be
due only to binary scattering events at discrete points in $s$, with all
particles moving along free-particle world lines between such binary
scatterings [\,The only way in which mean field effects enter the model
is through the fact that particles can be off-shell, thus acquiring
effective masses (cf.\ Sects.\ref{sec:ini_state},
\ref{sec:parton_scattering} below).\,].

Between the discontinuous binary interactions, the world lines of all
particles are given by the free Hamiltonian:
\begin{eqnarray*}
   p_i(s) &=& \text{const} \\
   x_i(s) &=& \frac{p_i}{m_i}\cdot (s-s_0) + x_i(s_0) \ic
\end{eqnarray*}
where $s_0$ is the last $s$ at which particle $i$ underwent an
interaction. Note that as a consequence of these world lines the
3-velocities are given by
\[ \frac{d\vec{r}_i}{dt_i} = \frac{d\vec{r}_i}{ds}\cdot\frac{ds}{dt_i}
                           = \frac{\vec{p}_i}{m_i}\cdot\frac{m_i}{E_i}
                                                                  \ic \]
as should be.

For a given $s$, the square of the \emph{Poincar\'e-invariant}
\text{4-distance} $d_{ij}$ between particles $i$ and $j$ is defined to
be
\[ {d_{ij}}^2 := -\left(x_\mu -\frac{(xp)}{p^2}p_\mu\right)
                 \left(x^\mu -\frac{(xp)}{p^2}p^\mu\right) \ic     \]
where $x$ and $p$ are the relative 4-distance and the total 4-momentum
of particles $i$ and $j$:
\[ x = x_i - x_j , \; \;  p = p_i + p_j \ip             \]
This is, of course, just a Poincar\'e-invariant way to write the
3-distance in their center-of-momentum frame (i.e\ their impact
parameter):
$|d_{ij}| = \left|\vec x_i - \vec x_j\right|_{\text{(CMS)}}$ .

Whenever two particles approach each other to a \text{4-distance} within
\begin{equation} \label{eq:dij}
    d_{ij} < \sqrt{\frac{\sigma_{\text{tot}}}{\pi}} \ic
\end{equation}
there will be a binary interaction between them. At that point in $s$,
their momenta change discontinuously. What interaction takes place and
how the interaction-distance (i.e.\ $\sigma_{\text{tot}}$) is determined
depends on the particular model.

The basic algorithm of the dynamics of our parton cascade is thus as
follows:
\begin{enumerate}
  \item in an initialization procedure the colliding nuclei are
        described in terms of a certain number of partons with initial
        phase space coordinates (cf.\ Sect.\ \ref{sec:ini_state}),
  \item the partons propagate through phase space until the first two of
        them are about to interact (at a given $s$),
  \item for that pair, the type of interaction is determined (cf.\
        Sect.\ \ref{sec:parton_scattering}). Additional partons may be
        produced in the process (cf.\ Sect.\ \ref{sec:evolution}).

        Due to this interaction, the interacting partons acquire new
        4-momenta,
  \item all partons continue to propagate freely until the next earliest
        $s$ for which another pair is up for an interaction,
  \item steps 3 and 4 are iterated until all partons move away from one
        another. The cascade then ends.
\end{enumerate}

Since the world-lines are parametrized by the
\emph{Poin\-car\'e-invariant} parameter $s$, the ordering of binary
interactions (determined by the sequence of parameters $s_{ij} < s_{kl}
< s_{mn} \dots$) is independent of the observer frame of reference in
which the cascade is run. This is in sharp contrast to the violation of
Poincar\'e-covariance generally encountered in cascade models involving
actions at a distance \cite{Kod84}.
\section{The initial state}            \label{sec:ini_state}

In contrast to most analytical transport models of ultrarelativistic
nucleus-nucleus collisions, we do not assume an equilibrium initial
state. Rather, we start out by first describing both colliding nuclei as
ground state configurations (cf.\ Sect.\ \ref{sec:distr_nuc}), which are
then boosted according to the kinematics of the particular reaction we
want to simulate. In a second step of the initialization (but before any
collisions occur) the individual nucleons are described in terms of a
set of (classical) `partons' (cf.\ Sect.\ \ref{sec:distr_part}). The
nucleus-nucleus collision is then modelled as a sequence of partonic
interactions, i.e.\ we do not allow any initial nucleon-nucleon
collisions; although some such initial hadronic interactions will
certainly occur, we believe them to be unimportant in the energy range
of interest (RHIC energies).

\subsection{Distribution of the nucleons}
\label{sec:distr_nuc}

The nucleons in each of the two nuclei are initially assigned random
positions and momenta. In the rest frame of the nucleus, the
distributions are spherically symmetric while the radial distributions
are taken to be of Saxon-Woods form:
\[ w(r) \sim \left(e^\frac{r-R_0}{\Delta r}+1\right)^{-1}
         \text{\qquad (positions)}  \ic           \]
with $R_0 = 1.07 \cdot A^{1/3}$ fm, $\Delta r = 0.5$ fm,
\[ w(p) \sim \left(e^\frac{p-p_F}{\Delta p}+1\right)^{-1}
         \text{\qquad (momenta)  }  \ic           \]
with $p_F = \frac{\hbar}{r_0} \sqrt[3]{\frac{9}{4}\pi } \, \approx \,
            0.316$ GeV/c, $\Delta p = 0.03$ GeV/c
[\,For simplicity's sake, we use identical distributions for protons and
neutrons.\,]. The reason for using smeared-out distributions both in
positions and momenta in the nuclear ground-state configurations is that
this accounts (to a certain degree) for the uncertainty principle in
this classical phase space description; it has nothing to do with finite
temperature. For the same reason, we enforce a minimum spatial distance
between any 2 nucleons of 0.8~fm. The zeroth (time) components of the
position 4-vectors of the nucleons are irrelevant at this point because
they are set on the parton level (cf.\ Sect.\ \ref{sec:distr_part}). The
nucleon energies are then fixed so that every nucleon is on its mass
shell:
\[ {p^0}_i = {E^0}_i = \sqrt{\vec{{p}_i}^2 +{M_N}^2} \ic
             \;\; (i= 1, \dots , A) \ic \]
an finally both nuclei are boosted to the desired frame of reference
(depending on the particular reaction simulated).

\subsection{Distribution of the partons}          \label{sec:distr_part}

In a next step, each nucleon is resolved into initial partons. The
longitudinal momenta (and flavors
\footnote{In this paper, the term ``flavor'' is used to denote either
          a quark of given flavor in the usual sense, or a gluon.
         })
of the generated partons are chosen randomly with distributions
corresponding to the experimental proton/neutron structure functions
$F_2(x,Q^2)$ in the form of the `GRV94LO' parametrization
\cite{Glu90,Glu95}, until a total number of partons $N$ is reached so
that in a given nucleon
\[ \sum_i^N x_i := \sum_i^N \frac{{p_z}_i}{P_z} \approx 1   \]
is satisfied.

Since the parton distribution functions peak at $x=0$, we need to use a
cutoff in $x$: $x_i\geq x_{\text{min}}$, which we choose inversely
proportional to the nucleon momentum $P_z$. This cutoff is also in
keeping with our focus is on hard partons.

Note that we need to specify an initial resolution scale ${Q_0}^2$ at
which we evaluate the structure-functions. The number of partons $N$
depends strongly on this scale, because the structure functions are
peaked at low $x$ for high $Q^2$. In contrast to the situation in, e.g.
deep inelastic lepton-proton scattering, in a heavy ion reaction there
is no clear definition for this scale, and we do not know it apriori. In
the context of the present section, viz.\ the construction of the
initial state, this initial resolution scale ${Q_0}^2$ is, therefore, an
arbitrarily chosen parameter (for the actual values used in the
numerical computations cf.\ Sect.~\ref{sec:results}). In Sect.\
\ref{sec:evolution}, we shall, however, discuss a parton evolution
mechanism which turns out to weaken the dependance of our final results
on ${Q_0}^2$ drastically.

As for the transverse momenta ${p_{\mathsf{T}}}_i$ of the partons, we take
these to be distributed radially symmetric about the nucleon momentum  $P_z$,
with a Gaussian distribution for their modulus:
\[   w(p_{\,\mathsf{T}}) \sim e^{-(p_{\,\mathsf{T}})^2/(2a^2)}  \ic   \]
with $a=0.3\;\text{ GeV}/c$ (cf.~\cite{Bel89}).

Because of the radial symmetry of the $p_{\,\mathsf{T}}$ distribution,
$\sum_i {p_{\,\mathsf{T}}}_i \approx 0$, so that for the partons in each
nucleon we have
\[ \sum_i {\vec{p}}_i \approx (0,0,P_z) \ip \]
As a consequence, this resolution into partons leaves the
\emph{nucleons on the mass shell}:
\[ \left( \sum_{i=1}^{N} {p_i}^\mu \right)
   \left( \sum_{i=1}^{N} {p_i}_\mu \right)
   \approx {M_{\text{nucleon}}}^2  \ip \]

The final momentum component to be fixed is the energy of each parton.
As was pointed out in Sect.\ \ref{sec:dynamic}, in PCD the energy of a
particle is not determined by a mass-shell condition, but is an
independent dynamic variable. This feature of the dynamics of our model
allows, in a very natural way, to use effective parton masses
(`virtualities') to satisfy other constraints given by the physics of
the initial state to be constructed. One such constraint is that
initially (before the first interactions between partons from one of the
colliding nuclei and partons from the other) all partons should remain
essentially confined within their repective nucleons. Without
constraining their velocities explicitly in some way, the partons would
spread out over the whole phase-space very quickly after initialization.

One possibility for a `confinement constraint' is to require the parton
longitudinal velocities to equal the velocity of the resolved nucleon:
\begin{equation}
   \label{eq:condition}
  {\beta _z}^{\text{parton}} = |\vec{\beta}^{\,\text{nucleon}}| \ip
\end{equation}
Since the parton transverse momenta are small compared to the
longitudinal momenta, this guarantees that the partons of one nucleon
move together for some time initially. We thus demand explicitly that
\[ {\beta _z}^{\text{parton}}
     = \frac{x_i P_z}{\sqrt{(x_i P_z)^2 + {{p_{\,\mathsf{T}}}_i}^2
                      +{\mu_i}^2 }}
     \stackrel{!}{=} \frac{P_z}{\sqrt{{P_z}^2+M^2}}  \ic \]
where the effective mass of parton $i$ is denoted by $\mu_i$, from which
one obtains
\begin{equation}  \label{eq:eff-mass}
    {\mu_i}^2 = {x_i}^2 M^2 - {{p_{\,\mathsf{T}}}_i}^2 \ip
\end{equation}
The fact that the confined partons acquire effective masses in this way
fits well into the framework of PCD, where interacting particles are
off-shell. There is, however, a technicality involved in this method of
modelling initial confinement: from (\ref{eq:eff-mass}) it can be seen
directly that whenever the transverse momentum of a parton is
sufficiently large, ${\mu_i}^2$ becomes negative, i.e.\ the parton
becomes superluminal. In order to avoid such particles, we reject
transverse momenta that lead to $\beta>1$. This means that our
transverse momentum distribution is no longer an exact Gaussian, but
somewhat narrower.

Finally, the spatial coordinates of the partons are chosen randomly with
a spherical distribution (centered at the spatial coordinate of the
nucleon) according to
\[  w(r) \sim e^{-4.33\;\text{\tiny fm}^{-1} \; r} \ic \]
in accordance with nuclear form factor data \cite{Hof58}. These
spherical distributions are then Lorentz-contracted along the beam axis,
and the $\text{zero}^{\text{th}}$ components of the 4-vectors $r^\mu$
are set to zero.

\subsection{Distributed Lorentz Contraction (DLC)?} \label{sec:dlc}

It has been proposed in the literature \cite{Bjo75,Gei95} to use a
``Distributed Lorentz Contraction (DLC)'' for the partons in order to
enlarge the longitudinal extension of a nucleus and thus to enhance the
chances for scatterings in a cascade model. The physical picture behind
such an idea can be described roughly as follows. While the longitudinal
extension of the \emph{valence quarks} in a fast-moving nucleon does
indeed look Lorentz-contracted to a stationary observer in the usual
way:
\[ (\Delta z)_{v} \approx \frac{2R_0}{\gamma}  \ic \]
the same is \emph{not true} for the \emph{sea-quarks and gluons}.
Rather, the longitudinal extension of sea-quarks and gluons should
always be at least of the order of $\sim 1$ fm; in a qualitative way one
argues that due to the uncertainty principle
\[ (\Delta z)_{g,s} \approx \frac{1}{p_z} = \frac{1}{xP} \ic \]
so that for smaller $x$ one has a larger longitudinal extension.

The practical argument usually given for using a DLC in a parton cascade
model at ultrarelativistic energies is that without it the extreme
Lorentz contraction of the colliding nuclei would simply not provide
enough time for their partons to interact sufficiently. With DLC, the
probablity for multiple scatterings, in particular, increases, thus
enhancing the possiblity of obtaining high temperatures and densities.

This, however, is an argument that makes sense only if one looks at the
physics from a given observer frame. While it is true that nuclei become
pancakes when we \emph{look} at them from a reference frame with high
relative velocity, the number of scatterings should \emph{not depend on
the reference frame} at all.

Thus, both of the above arguments for a DLC are extraneous to a formally
covariant formulation, and in our covariant parton cascade we do not
need such an ad-hoc prescription to enlarge the number of scatterings
[cf.\ Sect.\ \ref{sec:results}]. Indeed we have \emph{not employed} any
DLC-like prescription in producing the initial state: using the
invariant distance $d_{ij}$ to determine binary interactions guarantees
that the desired effects for which DLC was proposed are correctly and
automatically included in our code.
\section{Parton Scattering}            \label{sec:parton_scattering}

As detailed in Sect.\ \ref{sec:dynamic}, the cascade algorithm needs two
inputs from binary scattering:
\begin{enumerate}
  \item[(i)] a parton total cross section $\sigma_{\text{tot}}$, which
       determines whether a given pair of partons will come within an
       (invariant) interaction distance given by (\ref{eq:dij}),
  \item[(ii)] a differential cross section
       $\frac{d\sigma}{d\theta}(ab \rightarrow cd)$,
       which determines the details of an actual binary scattering.
\end{enumerate}

In the present section we describe how we obtain and use these cross
sections, i.e.\ we discuss ($2\rightarrow 2$)--scattering only. As
mentioned above, our model allows for the creation of additional partons
during a ($2\rightarrow 2$)--scattering. The mechanism for these
($2\rightarrow n$)--interactions will be discussed in
Sect.\ \ref{sec:evolution}.

\subsection{Notation (kinematics)}

Our notation is as follows. We use the Mandelstam variables
\[ s:= (p_1 +p_2)^2 \ic \text{ \quad }
   t:= (p_1 -p_3)^2 \ic \text{ \quad }
   u:= (p_1 -p_4)^2 \ic   \]
where the incoming particles are $p_1,p_2$, the outgoing ones $p_3,p_4$,
as usual. The momentum transfer $t$ is kinematically restricted to the
intervall $t_\pi \le t \le t_0$, where the subscripts ($0,\pi$) denote
the corresponding scattering angle in the CMS. Whereas for equal-mass
particles we have the familiar relation $-[s -(2m)^2] \le t \le 0$, for
four different masses one has
\begin{eqnarray*}
 t_0/t_\pi &=& \frac{1}{4s}
               \left\{{m_1}^2 -{m_2}^2 +{m_3}^2 -{m_4}^2\right\}^2
                                                           \\[1ex]
           && -\frac{1}{4s}
                \left\{\lambda(s,{m_1}^2,{m_2}^2)
                   \mp \lambda(s,{m_3}^2,{m_4}^2) \right\}^2 \ic \
\end{eqnarray*}
with the usual abbreviation
\[ \lambda(a,b,c):= \sqrt{a^2 +b^2 +c^2 -2ab -2ac -2bc} \ip  \]

The CMS scattering angle $\theta$ is given by
\[  \sfrac{1}{2}(1 - \cos\theta) = \sin^2\sfrac{\theta}{2}
         = \frac{t_0 - t}{t_0 - t_\pi}  \ip  \]

In Sect.\ \ref{sec:ini_state} we have set the initialized partons
off-shell: ${p_i}^2 = {\mu_i}^2 \neq {m_i}^2$, where $m_i$ denotes the
current mass and $\mu_i$ the effective mass of parton $i$. This
procedure was used there to model the initial confinement of quarks in
the nucleons of the colliding nuclei; it is out of place in the context
of perturbative QCD which we will be discussing here. In calculating the
Mandelstam variables for a particular ($2\rightarrow 2$)--interaction,
we therefore reset the two incoming partons to be on-shell by adjusting
the zeroth component of their momenta according to
\[ ({p_i})^0 := \sqrt{{\vec{p}_i\,}^2 +{m_i}^2} \ic  \]
and use the current masses for the outgoing partons as well
[\,cf.\ however, Sect.\ \ref{sec:virtualities}\,].

In some cases, in order to render the integrated cross sections finite,
it will be necessary to impose a cut $t_c >0$ on the kinematic limits of
the momentum transfer $t$, i.e.\ we restrict $t$
to $t_\pi \le t \le t_0 -t_c$, so that the integrated cross section for
a particular process will be
\begin{equation}        \label{eq:sigma}
   \sigma (s) = \int_{t_\pi}^{t_0-t_c} \frac{d\sigma}{dt}\,dt
\end{equation}
(in the case of identical particles, a similar cutoff needs to be
applied at $\theta=\pi$).

This cutoff procedure will be discussed in further detail below.

\subsection{Matrix elements}
The relevant ($2\rightarrow 2)$ matrix elements within perturbative QCD
in the tree-level approximation (cf.\ e.g.\
\cite{Bab77,Com77,Cut78,Glu78,Jon78,Com79,Isg79} are given below. They
take into account non-vanishing parton masses throughout (cf.\
\cite{Sve88,Nas89}). The results can be expressed by four functions
$G_0, \dots G_3$ of the Mandelstam variables (and the masses) of the
scattering particles:
\begin{eqnarray*}
   G_0(s,t,u)                                       
              &=& \left\langle\sfrac{9}{2}\right\rangle\,
                  \left(3 -\frac{ut}{s^2} -\frac{su}{t^2}
                 -\frac{ts}{u^2} \right)
\end{eqnarray*}
\begin{eqnarray*}
  \lefteqn{G_1(s,t,u;m, m') =                 
                   \left\langle\sfrac{2}{9}\right\rangle\,
                   \frac{2}{t^2}\left\{
                      \left[s - (m^2+{m'}^2)\right]^2 \right.}  \\
              &&   \left. +t\left[(m^2+{m'}^2 -u)^2
                          +2(m^2+{m'}^2)\right]
                                                \right\}
\end{eqnarray*}
\begin{eqnarray*}
   G_2(s,t,u;m)                                     
              &=& -\left\langle -\sfrac{2}{27}\right\rangle\,
                   \frac{4}{tu} \left(s -2m^2\right)
                                \left(s -6m^2\right)
\end{eqnarray*}
\begin{eqnarray*}
  \lefteqn{G_3 (s,t,u;m) =                    
                  \left\langle\sfrac{3}{16}\right\rangle
                  \frac{4(m^2-t)(m^2-u)}{s^2}           }         \\
              && +\left\langle\sfrac{1}{12}\right\rangle
                  \frac{2(m^2-u)(m^2-t) -4m^2(m^2+t)}{(t-m^2)^2}   \\
              && +\left\langle\sfrac{1}{12}\right\rangle
                  \frac{2(m^2-t)(m^2-u) -4m^2(m^2+u)}{(u-m^2)^2}   \\
              && +\left\langle-\sfrac{1}{96}\right\rangle
                  \frac{-4\left[2m^2(s-2m^2) +(m^2-t)(m^2-u) \right]}%
                       {(m^2-t)(m^2-u)}                            \\
              && +\left\langle\sfrac{3}{32}\right\rangle
                  \frac{4\left[t(s+t) -m^4 \right]}{s(m^2-t)}      \\
              && +\left\langle\sfrac{3}{32}\right\rangle
                  \frac{4\left[u(s+u) -m^4 \right]}{s(m^2-u)}      \ip
\end{eqnarray*}
In the above expressions, the numerical factors given in brackets are
due to the various color averages.

All relevant parton matrix elements $|{\cal M}^2|$  (with the strong
coupling constant $\alpha_s=g^2/4\pi$ factored out) can be expressed in
terms of these functions. Processes with different incoming and outgoing
particles, but the same topology of the Feynman diagrams are related to
one another by crossing, i.e.\ in our case by the interchange of the
appropriate Mandelstam variables in the functions $G_i$. The resulting
relations between the matrix elements and the functions $G_i$ are
given in Tab.\ \ref{tab:subprocesses}.

\subsection{The scattering process}
The matrix elements listed in Tab.\ \ref{tab:subprocesses} determine the
differential cross section for a particular process in the standard way:
\[  \frac{d\sigma}{dt} (ab \rightarrow cd)
    = \frac{|g^2{\cal M}|^2 }{64\pi |\vec p_1|^2 s}
  \approx \frac{\pi \alpha_s^2}{s^2}|{\cal M}|^2  \ip  \]
From this we obtain the total cross section to be used in (\ref{eq:dij})
by integrating and summing over all channels. To this end we first need
to compute the values of the integrated partial cross sections
$\sigma_i$ at the given CMS energy $s$ for all possible channels, e.g.\
$\sigma(u\bar{u}\rightarrow u\bar{u}), \sigma(u\bar{u}\rightarrow
d\bar{d}), \ldots ,\sigma(u\bar{u}\rightarrow gg)$ in the case of a
$u\bar{u}$ pair in the initial state. If the condition (\ref{eq:dij})
determines that a binary scattering is indeed to take place, the
specific process to actually occur is determined randomly, with weights
given by the relative sizes of the $\sigma_i$. We then similiarily
choose the momentum transfer $t$ (and thus the CMS scattering angle
$\theta$) by sampling the appropriate differential cross section
$d\sigma/dt$ (the CMS azimuth angle $\phi$ is, of course, chosen with an
isotropic distribution).

In all of the above considerations, we have dropped inelastic
($2\rightarrow 2$)--processes. Since all processes with partons of
different flavour in the initial and final state have a typical
$s$--channel behavior at high energies, the contribution from the
inelastic cross sections is insignificant.

Several points in the procedures described still need further
clarification. We now discuss these in order.

\paragraph*{$\mathbf{t}$ \textbf{cutoff .}}
Some of the scattering matrix elements (cf.\
Tab.\ \ref{tab:subprocesses}) have the typical Rutherford singularity in
the forward direction (in the case of identical particles in the
backward direction also). In order to obtain finite integrated cross
sections, we impose a kinematic cut $t_c$ on the momentum transfer $t$.

To justify this procedure, let us consider for a moment the processes of
soft radiation in the initial and final state. These processes are
dominant around the divergences in question. Including this soft
radiation would render the poles finite or at least only logarithmically
divergent, the divergences thus turn out to be a consequence of the
perturbative approximation. So if we want to remain consistent in
considering only `hard' processes in the context of the present section,
we should omit contributions from the vicinity of the poles altogether.
The soft region will be `resummed' later, when we include an evolution
scheme based on the DGLAP equations (cf.\ Sect.\ \ref{sec:evolution}).

For the method of cutoff, we have an alternative choice between two
physically different possibilities. In the first $t_c$ is basically
constant. In the second $t_c$ is determined by the CMS energy $s$ of the
particular interaction, corresponding e.g.\ to a minimum scattering
angle in the CMS of the two scattering particles. Although the latter
option may seem more intuitive, the implicit $s$--dependence of the
resulting cutoff leads to singular behavior of the total cross sections
close to the kinematic threshold in $s$. The former option will keep the
cross sections smooth also in the region close to the threshold and is
therefore preferable. In the numerical computations we have used $t_c=
{Q_{\text{min}}}^2$ (cf.\ below).

\paragraph*{$\mathbf{\bbox{\alpha}_s(Q^2)}$ \textbf{and}
            $\mathbf{Q^2}$ \textbf{cutoff}.}
We employ renormalization group--improved perturbation theory, e.g.\ we
use a running coupling $\alpha_s = \alpha_s(Q^2)$, thereby including
some higher order perturbative effects in a qualitative way. The `scale'
$Q^2$ may in general be a function of all of the Mandelstam variables
for the particular process. The choice of $Q^2$ is not obvious, since in
a collision of many hadrons there is no external scale that determines
$Q^2$, as is the case e.g.\ in deep inelastic scattering, and several
possibilities have been discussed in the literature \cite{Com79,Gst90}.
For practical reasons, we have simply used $Q^2 = s -{m_1}^2 -{m_2}^2$
in our model, thus neglecting a possible (logarithmic) dependence of
$\alpha_s(Q^2)$ on the momentum transfer $t$.

A further point is that the whole picture of parton binary scattering
described and computed with perturbative QCD (`hard scattering')
implies, of course, a small value of $\alpha_s$. Consequently, for such
a description to be consistent, the value of $Q^2$ should not fall below
some given value. In implementing this point in the cascade algorithm,
we cut off the allowed range of $Q^2$, i.e.\ we do not allow partons to
scatter at all if $Q^2 < {Q_{\text{min}}}^2$, using the method proposed
in \cite{Abo92} (for a detailed discussion of this issue cf.\
\cite{Sat00}. The actual values of ${Q_{\text{min}}}^2$ used in the
numerical computations are given in Tab.\ \ref{tab:parameters}).

Both of these choices: $Q^2 = s -{m_1}^2 -{m_2}^2$, and no scattering
for $Q^2 < {Q_{\text{min}}}^2$, are to some degree arbitrary. However,
this arbitrariness is again mitigated by the parton evolution mechanism
described in Sect.\ \ref{sec:evolution}.

\subsection{Virtualities}           \label{sec:virtualities}
The cascade approach, which assumes its constituents to be moving freely
between instantaneous scatterings, is, of course, a very drastically
simplified model of a system of strongly interacting particles. It is
one of the virtues of the PCD dynamics that it allows to model mean
field effects by allowing particles to be off shell (`virtual') in a
natural way. In the initialization of our cascade, we have used this
feature to model the confinement of quarks in the initial nucleons
cf.\ Sect.\ \ref{sec:ini_state}).

We can make use of the same feature again to model some of the effects
of the nuclear medium (the QGP?) on the motion of partons during the
nuclear reaction. As before, instead of introducing mean-field effects
via a PCD quasipotential, we choose to introduce parton virtualities
directly in every ($2\rightarrow 2$)--interaction.

The specific implementation of such parton virtualities is, of course,
restricted by the requirement of 4--momentum conservation, but there
still are several possibilities [\,in all cases studied, we have
included parton virtualities by adjusting the zeroth components
(energies) of the outgoing partons in quite an analogous way in which we
have set the ingoing partons on-shell before scattering, viz.\ we keep
the spatial components of their momenta, $\vec p_3$ and $\vec p_4$,
fixed at the values determined in the scattering \emph{using on-shell}
(current) masses, and then adjust $(p_3)^0$ and $(p_4)^0$ subject to
energy conservation $(p_3)^0 +(p_4)^0 =(p_1)^0 +(p_2)^0$\,].

In detail we have investigated the following schemes: after a binary
scattering event
\begin{enumerate}
  \item one of the two outgoing partons is left in the state determined
        by the scattering process, i.e.\ it remains on-shell. The other
        outgoing parton attains an effective mass, which is uniquely
        determined by energy conservation in the procedure described
        above.

        In deciding which parton to leave on-shell, we can either choose
        at random (unbiased choice), or we can select the parton with
        the larger tranverse momentum. A qualitative argument for the
        latter would be that the parton with the larger tranverse
        momentum leaves the dense zone of the nuclear medium sooner and
        is thus less subject to the effects of the medium (which is just
        what is being modelled by the virtualities).
  \item the effective masses of both outgoing partons are obtained by
        \emph{adding the same amount} of virtuality to their current
        masses, subject to energy conservation. The virtualities are
        again determined uniquely.
  \item the effective masses are obtained by \emph{multiplying by the
        same factor} the CMS energies of both outgoing partons. It
        should be noted that this scheme is not covariant, since it uses
        the CMS in an essential way.
  \item we do not add any virtualities at all, i.e.\ the outgoing
        partons are left on-shell.
\end{enumerate}
Although the effects of the different schemes on, e.g., the total number
of scatterings suffered by a parton in the course of a nuclear collision
are not negligible, the first of these schemes has turned out to be the
most viable one, and all numerical results given in
Sect.\ \ref{sec:results} were obtained with it. It also seems to be the
choice best motivated by a physical argument.

Finally, we wish to point out that we have \emph{not modified} the
leading-order parton cross sections by a ``K-factor'' (i.e.\ $K=1$ in
PCPC throughout, whereas comparable models usually use $K=2$ to $K=3$).
In the context of our somewhat different approach to higher-order
corrections as described in this and the following Section (Sect.\
\ref{sec:evolution}), the reasons for introducing a K-factor do not seem
clear (for a recent discussion of the K-factor and its relevance in
parton cascades cf.\ \cite{Bas99}).

We close this section by summarizing all the options introduced in the
implementation of the partonic scattering:
\begin{itemize}
   \item the method of cutting off the poles of the differential cross
         sections (including the choice of cutoff parameters $t_c$)
   \item the choice of the argument $Q^2$ in the running coupling
         constant $\alpha_s(Q^2)$
   \item the introduction of a minimal scale ${Q_{\text{min}}}^2$ below
         which any `hard' interaction will be excluded
   \item the method of assigning new virtualities to
         the particles after a binary collision.
\end{itemize}
\section{Parton Evolution}             \label{sec:evolution}

As was mentioned before, we also allow for $(2 \rightarrow n)$ processes
in our model. In the present section we describe the emission of soft
parton radiation before a `hard' parton scattering takes place.

Let us recall (cf.\ Sect.\ \ref{sec:ini_state}) that our cascade starts
with an initial ensemble of partons which are resolved at a rather small
scale (typically ${Q_0}^2 \approx 10\text{ GeV}^2$). We now interpret
these initial partons as ``\emph{pre-partons}'', to be resolved further
in a $(2 \rightarrow 2)$--scattering process, with a scale
${Q_h}^2>{Q_0}^2$, by means of the DGLAP parton evolution.

In order to employ this mechanism for soft parton radiation, we would
need do know the longitudinal momentum fraction of the parton to be
radiated, i.e.\ the scale ${Q_h}^2$, whereas at this point of our
algorithm we know only the momentum fraction $x=\frac{p_a}{P}$ of the
whole pre-parton $a$. We shall deal with this problem in part C of this
section.

During the interaction we fix the structure of the pre-parton, i.e.\ we
first determine the number of soft partons radiated (cf.\ part B of this
section), and then fix their properties , viz.\ (i) the flavors they
carry, (ii) their (off-shell) 4-momenta and, as our model is a
space-time description, also (iii) their 4-positions (cf.\ part C).

\subsection{The model for soft partons}
\label{sec:soft_parton}

\begin{figure}
     \centerline{\includegraphics[keepaspectratio,width=75mm]%
                 {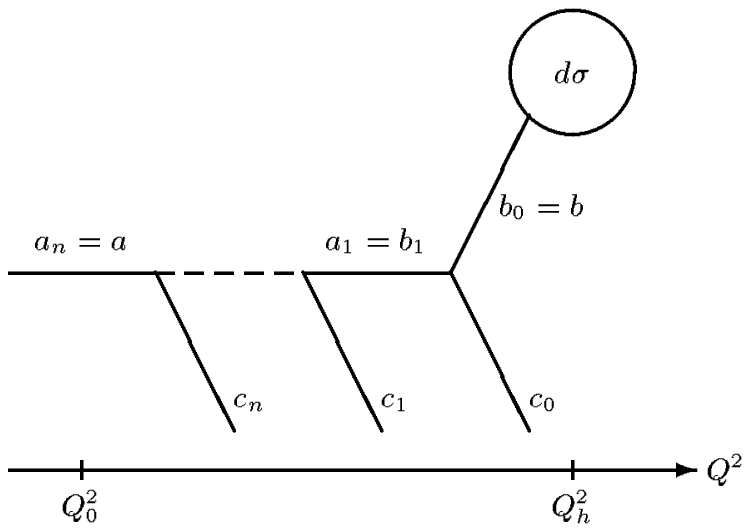}
                }
     \caption{The branching chain for a scattering process. The incoming
              pre-parton $a$ radiates secondary partons $c_i$ with
              increasing scale $Q^2$ and thus is resolved into its
              substructure.
             }               \label{fig:chain}
     \vspace*{1ex}
\end{figure}

 We follow the parton evolution by constructing a chain of successive
branchings for each colliding pre-parton $a$, as depicted in
Fig.\ref{fig:chain}. To this end, we make use of the \emph{Sudakov form
factor} \cite{Sud56}, which is essentially an integration of the DGLAP
evolution equations \cite{Dok77,Gri72,Lip75,Alt77}:
\begin{equation}                       \label{eq:sudakov}
  S(x_b, {Q_h}^2; Q^2) =
     \exp\! \left[\! -\!\! \int\limits_{Q^2}^{{Q_h}^2} \!\!
     \frac{d{Q'}^2}{{Q'}^2}
     \frac{\alpha_s({Q'}^2)}{2\pi}
     \sum_{a,c} W_{a,bc}({Q'}^2) \right]
\end{equation}
with
\begin{equation}                       \label{eq:type}
W_{a,bc}({Q'}^2) := \int\limits_{x_b}^1 \frac{dz}{z}
                  \frac{f_a(x_b/z, {Q'}^2)}{f_b(x_b, {Q'}^2)}
                  P_{a\rightarrow bc}(z)  \ic
\end{equation}
where $P_{a\rightarrow bc}(z)$ are the Altarelli-Parisi splitting
functions \cite{Alt77} [cf.\ Tab.\ref{tab:APsplitting}], and the
$f_a(x,Q^2)$ are the nucleon structure functions for partons with flavor
$a$ (we use the parametrization \cite{Glu90}).

${Q_h}^2$ is the scale of the  hard scattering. In general this can be a
function of all the kinematical invariants of the $(2 \rightarrow
2)$--scattering; for simplicity we choose it in line with the
corresponding choice in Sect.\ \ref{sec:parton_scattering}, viz.
\begin{equation}                    \label{eq:hardscale}
  {Q_h}^2 = (p_a +p_{a'})^2 -m_a^2 -m_{a'}^2
          = s -m_a^2 -m_{a'}^2               \ic
\end{equation}
here $p_a$ and $p_{a'}$ are the 4-momenta of the two incoming
pre-partons $a$ and $a'$, respectively.

As suggested by Fig.\ref{fig:chain}, the Sudakov form factor describes
a summation of all the soft parton modes which we are excluding in the
description of hard scattering (cf.\ Sect.\ \ref{sec:parton_scattering}).
More precisely, it is a summation of all diagrams similar to
Fig.\ref{fig:chain}, with the summation including a sum over the number
of vertices. In our algorithm, in any specific interaction we determine
a definite number of vertices, and thus generate a definite number of
soft partons with explicit flavors and momenta, as will be explained
presently.

\subsection{The branching chain} \label{sec:chain}

To construct the branching chain, we use a ``backward evolution''
algorithm \cite{Sjo85,Ben86} to follow the parton from the resolution
scale ${Q_h}^2$ back to the initial resolution ${Q_0}^2$. This algorithm
was extended to multiple parton interactions in \cite{Gei92,Gei95}.

The Sudakov form factor is interpreted as the probability that a parton
that is resolved at a scale ${Q_h}^2$ will be the same all the way down
to scale $Q^2<{Q_h}^2$. In other words, we choose a value $Q^2<{Q_h}^2$
according to the probability distribution given by Eq.\
(\ref{eq:sudakov}), and interpret it as the scale where the previous
branching in the chain occurs. At this scale we assign flavors and
momenta to the parton and its secondary parton. We continue to find the
next scale for a further branching. The algorithm terminates when the
scale reaches the initial value ${Q_0}^2$.

In principle a successive resolution by single branchings should produce
a branching \emph{tree} (branching of partons $c$ as well, cf.\
Fig.\ref{fig:chain}). For simplicity we restrict ourselves to branching
\emph{chains}, as illustrated in Fig.\ref{fig:chain}. Thus, our
``backward evolution'' algorithm proceedes explicitly in four steps:
\begin{enumerate}
   \item determine the scale ${Q_i}^2 < {Q_h}^2$ at which a branching
         of parton $a_{i+1}$ into partons $c_i$ and $a_i$ occurs,
   \item assign flavors to partons $c_i$ and $a_{i+1}$,
   \item assign the other properties (momenta, virtualities, and
         positions) to partons $b_i$ and $c_i$,
   \item replace ${Q_h}^2$ with ${Q_i}^2$, and iterate steps 1-3 until
         ${Q_i}^2 \le {Q_0}^2$.
\end{enumerate}
The number of successively obtained values $Q_i^2$ then gives us the
number of branchings and therefore the number of generated secondary
(soft) partons $c_i$.

\paragraph*{\textbf{Flavor.}}
The flavors of the partons at a particular point of a branching chain
are determined by the relevant vertex $a\rightarrow bc$. In the exponent
of Eq.\ (\ref{eq:sudakov}) there is a sum over all possible vertices
resulting in the final parton with flavor $b$. This sum reflects the
several branching channels and is restricted by flavor conservation: if,
e.g.,  parton $b$ is a quark, the associated parton $a$ is either a
quark of the same flavor or a gluon. The flavor of parton $c$ is then
also fixed completely by flavor conservation. The probability for each
allowed vertex $a\rightarrow bc$ is given by the relative weight of the
different terms in Eq.\ \ref{eq:type}.

For high momentum partons ($x\gtrsim 0.01$), the splitting is dominated
by soft gluon emission, and the sum in the exponent of
Eq.\ \ref{eq:sudakov} effectively reduces to a single term. For soft
partons ($x<0.01$) on the other hand, quark--antiquark production and
gluon emission are of the same order of magnitude, and the
quark--antiquark contributions to the sum cannot be neglected if one
wants to describe the production of heavy quarks (such as charmed
quarks) adequately. Our procedure guarantees that heavy partons are not
generated below their specific threshold scale (as given by the
parametrization of \cite{Glu90,Glu95}).

\subsection{The properties of soft partons}
\label{sec:parton_props}

In step 3 of the backward evolution algorithm, we need to assign
(i) 4-momenta, (ii) effective masses (virtualities), and (iii)
4-positions to the newly created partons in a vertex $a\rightarrow bc$.
In what follows, we describe the details of these assignments.

\paragraph*{\textbf{Longitudinal momentum fraction.}}
In determining the longitudinal momentum fraction $z=x_b/x_a$
(longitudinal with respect to the motion of parton $a$), we again refer
to the Sudakov form factor (Eq.\ \ref{eq:sudakov}). The integrand in
Eq.\ \ref{eq:type} represents the probability that a parton $a$ with
momentum fraction $x_a$ is resolved into a parton $b$ with momentum $x_b
\le x_a/z$. The momentum of parton $c$ is then determined by momentum
conservation.

Whenever there is a gluon in the final state, the splitting functions
are singular at $z=0$ and/or $z=1$. While the singularity at $z=0$ is
innocuous because $x_b>0$, we regularize the infrared divergence at
$z=1$ (soft gluon emission) by introducing a cutoff $z_{\text{max}}$,
thus restricting the integration interval in Eq.\ \ref{eq:type} to
$0<x_b<z<z_{\text{max}}<1$. We use
$z_{\text{max}} = x_b/\left(x_b+x_{\text{min}}\right)$, which allows for
gluons with momentum fraction $x_c \ge x_{\text{min}}$ only (for the
value of $x_{\text{min}}$ cf.\ Tab.\ \ref{tab:parameters}).

\paragraph*{\textbf{Transverse momenta and virtualities.}}
While the DGLAP parton evolution equations and Eq.\ \ref{eq:sudakov}
refer only to longitudinal momenta, in a nucleus-nucleus collision
transverse momenta play an important r{\^o}le. It is therefore
physically reasonable to supply the generated partons $b,c$ with some
transverse momentum $\vec{p}_{\mathbf{T}}$.

The parton momenta thus are
\begin{mathletters}                     \label{eq:kinematik}
\begin{eqnarray}
  p_a &=& \left(\sqrt{p^2 +{\mu_a}^2}                    \,;\;
          {\vec{0}}_{\mathbf{T}}, p       \right)                         \\
  p_b &=& \left(\sqrt{(zp)^2 +{p_{\,\mathsf{T}}}^2 +{\mu_b}^2}    \,;\;
          {\vec{p}}_{\mathbf{T}}, zp      \right)                         \\
  p_c &=& \left(\sqrt{(1-z)^2p^2 + {p_{\,\mathsf{T}}}^2 + {\mu_c}^2} \,;\;
          -{\vec{p}}_{\mathbf{T}}, (1-z)p \right)                \ic
\end{eqnarray}
\end{mathletters}
where ${\mu_i}^2 ={m_i}^2 -{q_i}^2,\quad i=(a,b,c)$, and the
difference ${q_i}^2$ between the current masses ${m_i}^2$ and effective
masses are the virtualities, as in
Sects.\ref{sec:ini_state},\ref{sec:parton_scattering}.

We now demand that the longitudinal velocities of the generated partons
$b$ and $c$ are the same as that of parton $a$, i.e.
\[                         
    \beta_{z_b} \stackrel{!}{=} \beta_{z_c}
                \stackrel{!}{=} |\vec{\beta}_a|    \ic   \]
and all particles have absolute velocities less than the speed of light.
This is analogous to our procedure in Sect.\ \ref{sec:ini_state}.
The first constraint leads to
\begin{eqnarray*}                       
  {\mu_b}^2 &:=& z^2{\mu_a}^2     -{p_{\,\mathsf{T}}}^2     \\
  {\mu_c}^2 &:=& (1-z)^2{\mu_a}^2 -{p_{\,\mathsf{T}}}^2     \ip
\end{eqnarray*}
Inserting these expressions for the effective masses in
Eqs.\ \ref{eq:kinematik} one finds that momentum is conserved in the
vertex in all four components, irrespective of the value of
${p_{\,\mathsf{T}}}^2$, so that we are indeed free to choose the transverse
momentum randomly. The virtualities are thus fixed in a purely kinematic
way. The second constraint, $\beta<c$, restricts the value of
${p_{\,\mathsf{T}}}^2$ to
\begin{equation}
  \label{eq:transmax}
  {p_{\,\mathsf{T}}}^2 \le \min \left\{
    \begin{array}{r} z^2{\mu_a}^2      \\
      (1-z)^2{\mu_a}^2
    \end{array} \right.                \ip
\end{equation}
We thus choose a ${\vec{p}}_{\,\mathsf{T}}$ randomly, with a distribution
that is radially symmetric about the axis given by $\vec{p}_a$, and
homogeneous up to the maximum value given by (\ref{eq:transmax}) . As
$0<z\le 1$, the invariant masses of the generated partons are always
\emph{less} than that of the parton $p_a$, so that the parton
virtualities increase along the branching chain from the pre--parton to
the scattering parton. This is an essential feature of whole idea of
parton evolution; it is interesting to note how naturally it is
accomodated in the PCD dynamical approach.

\paragraph*{\textbf{4-positions.}}
Finally, in accordance with the fact that the parton evolution occurs at
the same invariant $s$ parameter as the $(2 \rightarrow 2)$--scattering,
the 4-positions of all generated partons are set to that of the
pre--parton.

In summarizing, it is worthwhile to point out that only the last parton
(named $b_0=b$ in Fig.\ref{fig:chain}) scatters, whereas all others
leave the collision without further interaction. As noted in the
beginning of this section, before an interaction we know only the
momentum and flavor of the pre--parton, not that of the parton that
finally takes part in the $(2 \rightarrow 2)$--scattering. The situation
is complicated by the fact that it is the kinematics of the
$(2 \rightarrow 2)$--scattering event which tells us whether to start
the parton evolution algorithm in the first place. But as the branchings
are dominated by soft gluons, the longitudinal momenta of the
pre--parton and the colliding parton are nearly the same, and so it
seems justified to use $x_a$ (instead of $x_b$) in determining the total
cross section and thus the invariant $s$ parameter at which the
interaction is to take place.

We close this section by summarizing the options introduced in the
implementation of the parton evolution:
\begin{itemize}
   \item the form in which ${Q_h}^2$ depends on the kinematic variables
         (Eq.\ \ref{eq:hardscale})
   \item the restriction of the branching tree to a branching chain
   \item the choice of cutoff $z_{\text{max}}$ for regularizing the
         infrared divergence in Eq.\ \ref{eq:type}
   \item the choice of probability distribution for the transverse
         momenta of the radiated partons.
\end{itemize}
\section{Numerical Results}            \label{sec:results}
In this section we present numerical results of PCPC runs for various
nuclear reactions at various energies. While this paper primarily aims
at RHIC energies and heavy-ion reactions, and indeed, the parton cascade
approach itself is expected to be suited for this regime in particular
(and less so for, say, $p$$-$$\bar{p}$ reactions or heavy-ion physics at
SPS energies), there are as yet no experimental data available from
RHIC. In order to relate our results to experiment, we include some of
these other regimes as well.

In all of these cases, we essentially present final parton rapidity and
transverse momenta distributions. At this point we want to point out
once more that PCPC is a model for the dynamical evolution of
\emph{partons}; it does not deal at all with the hadronization of these
partons in the final (or at least late) stages of this evolution. As the
details of the physics of hadronization in heavy ion reactions are as
yet not fully understood (we can expect medium effects, in particular,
to play an increasingly important r{\^o}le), hadronization mechanisms in
parton cascade models for heavy ion reactions are at present
phenomenological at best, and often ad hoc. Nevertheless, we also
present some conclusions for \emph{hadron} rapidity and transverse
momenta distributions, deduced from our parton results with some very
simple assumptions; but we want the reader to keep in mind the
distinctly different character of these conclusions: they are not an
intrinsic part of our model.

\subsection{$p$$-$$\bar{p}$ reactions} \label{sec:ppbar}
This section presents the results of PCPC simulations of $p$$-$$\bar{p}$
reactions at various energies (parameter values cf.\ Table
\ref{tab:parameters}). They were obtained from a total of 5000
PCPC runs at each energy. On the average, 50 (at 200 GeV) to 170 (at
1800 GeV) partons per event were generated by the code; only about a
third of these have underwent a binary scattering or were generated with
the DGLAP mechanism described in Sect.\ \ref{sec:evolution}
(`participating partons'). Only these participating partons have been
included in the pseudorapidity distributions (and, indeed, in all
subsequent evaluations presented in this paper).

\begin{figure}
     \centerline{\includegraphics[keepaspectratio,width=75mm]%
                 {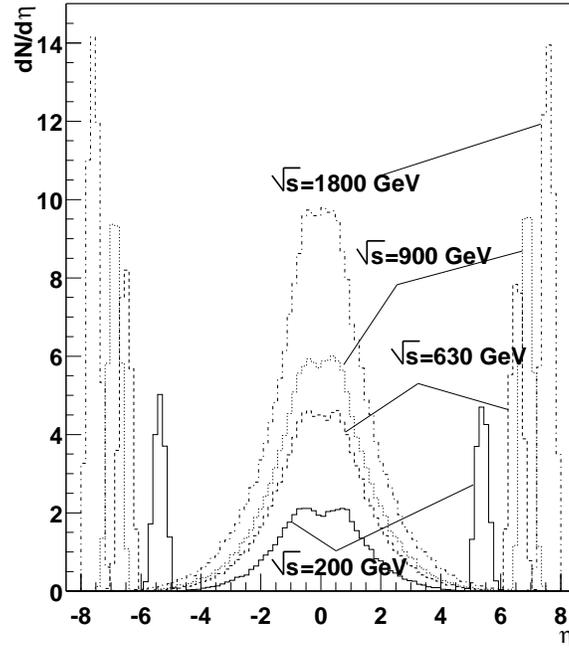}
                }
     \caption{Final pseudorapidity distributions (all partons)
              for a $p\bar{p}$ reaction at
              \protect\\
              $\sqrt{s}= 200,630,900
              \text{ and } 1800 \text{ GeV}$.
             }               \label{fig:ppbar-rap-all-200-1800GeV}
\end{figure}

The resulting pseudorapidity distributions of all participating partons
are given in Fig.\ref{fig:ppbar-rap-all-200-1800GeV}. Note that the
peaks at the beam rapidities do \emph{not} represent trivial spectator
partons, but are probably essentially DGLAP gluons.

\begin{figure}
     \centerline{\includegraphics[keepaspectratio,width=75mm]%
                 {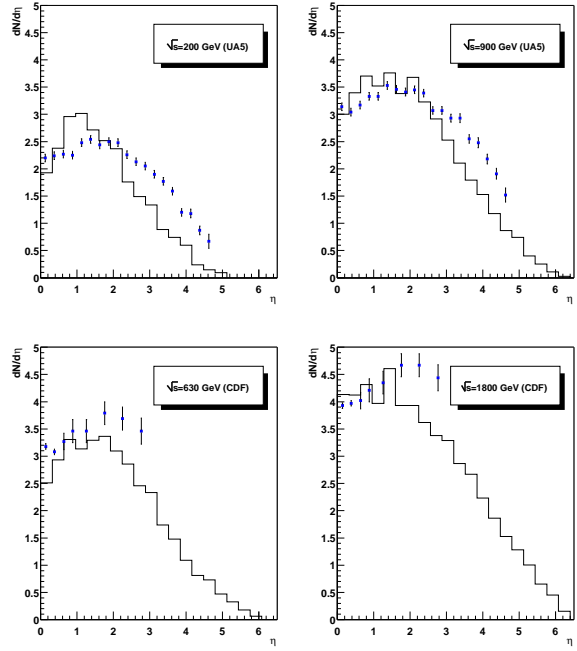}
                }
     \caption{Final pseudorapidity distributions (quarks only) for a
              $p$$-$$\bar{p}$ reaction at
              \protect\\
              $\sqrt{s}= 200,630,900 \text{ and } 1800 \text{ GeV}$. The
              histograms are the PCPC results. The data points are the
              experimental $\frac{dN_{ch}}{d\eta}$ (from
              \protect\cite{Aln86b} for 200 GeV, 900 GeV, and
              \protect\cite{Abe90} for 630 GeV, 1800 GeV, resp.).
             }               \label{fig:ppbar-rap-fit-200-1800GeV}
\end{figure}

Fig.\ref{fig:ppbar-rap-fit-200-1800GeV} shows the pseudorapidity
distributions of the participating quarks only. Since the number of
charged hadrons should be roughly proportional to the number of quarks,
we have included in Fig.\ref{fig:ppbar-rap-fit-200-1800GeV} some
experimental data for these reactions, as given in \cite{Aln86b,Abe90}.
Because the exact relation between quarks and charged hadrons depends on
a hadronization scheme, which is not part of our model, we have plotted
the quark distributions in Fig.\ref{fig:ppbar-rap-fit-200-1800GeV} with
an arbitrary scale (which, however, is the same for all 4 energies).

More interesting is the distribution of transverse momenta, as given in
Fig.\ref{fig:ppbar-pt-fit-200-1800GeV}. Since symmetry arguments suggest
that the distribution of baryon transverse momenta should be essentially
the same as that of the partons, these results can be compared directly
with experimental data. In Fig.\ref{fig:ppbar-pt-fit-200-1800GeV} we
have included the results of \cite{Alb90,Abe88}. Note that these plots
(both in the data and our simulations) involve a rapidity cut: only
particles with $|y|<1$ are included. The comparison shows that, apart
from the dips in the PCPC results at the lowest $p_{\mathsf{T}}$ (which are due
to the neglect of soft partons interactions), the agreement for
the transverse momenta is quite satisfactory; in fact, it improves with
increasing energy, pointing again to the decreasing importance of soft
parton interactions at higher energy.

\subsection{S$-$S and Pb$-$Pb at the SPS} \label{sec:SPS}
In this section we present the results of PCPC simulations for S$-$S and
Pb$-$Pb at the CERN SPS. The rapidity and $p_{\mathsf{T}}$ distributions we
present were obtained from 2000 PCPC runs (S$-$S reactions) and 500 runs
(Pb$-$Pb). On the average, 340 and 2360 partons per event were generated
by the code for S$-$S and Pb$-$Pb reactions, respectively. Of these,
29\% and 53\% were `participating partons'. Again, only the latter are
included in the distributions.

\begin{figure}
     \centerline{\hspace*{1.5pt}
                 \includegraphics[keepaspectratio,width=73.5mm]%
                 {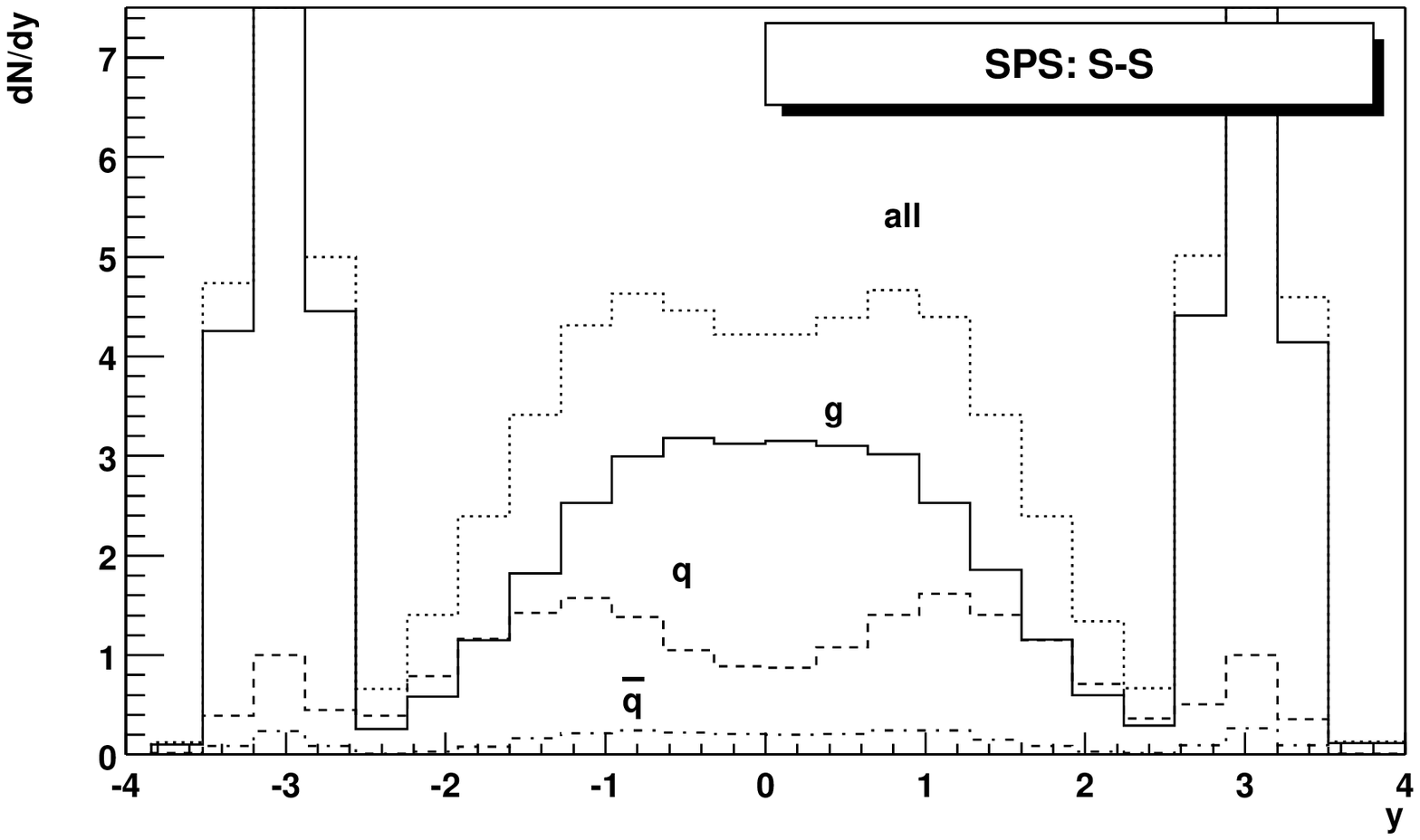}
                }
     \vspace{6pt}
     \centerline{\includegraphics[keepaspectratio,width=75mm]%
                 {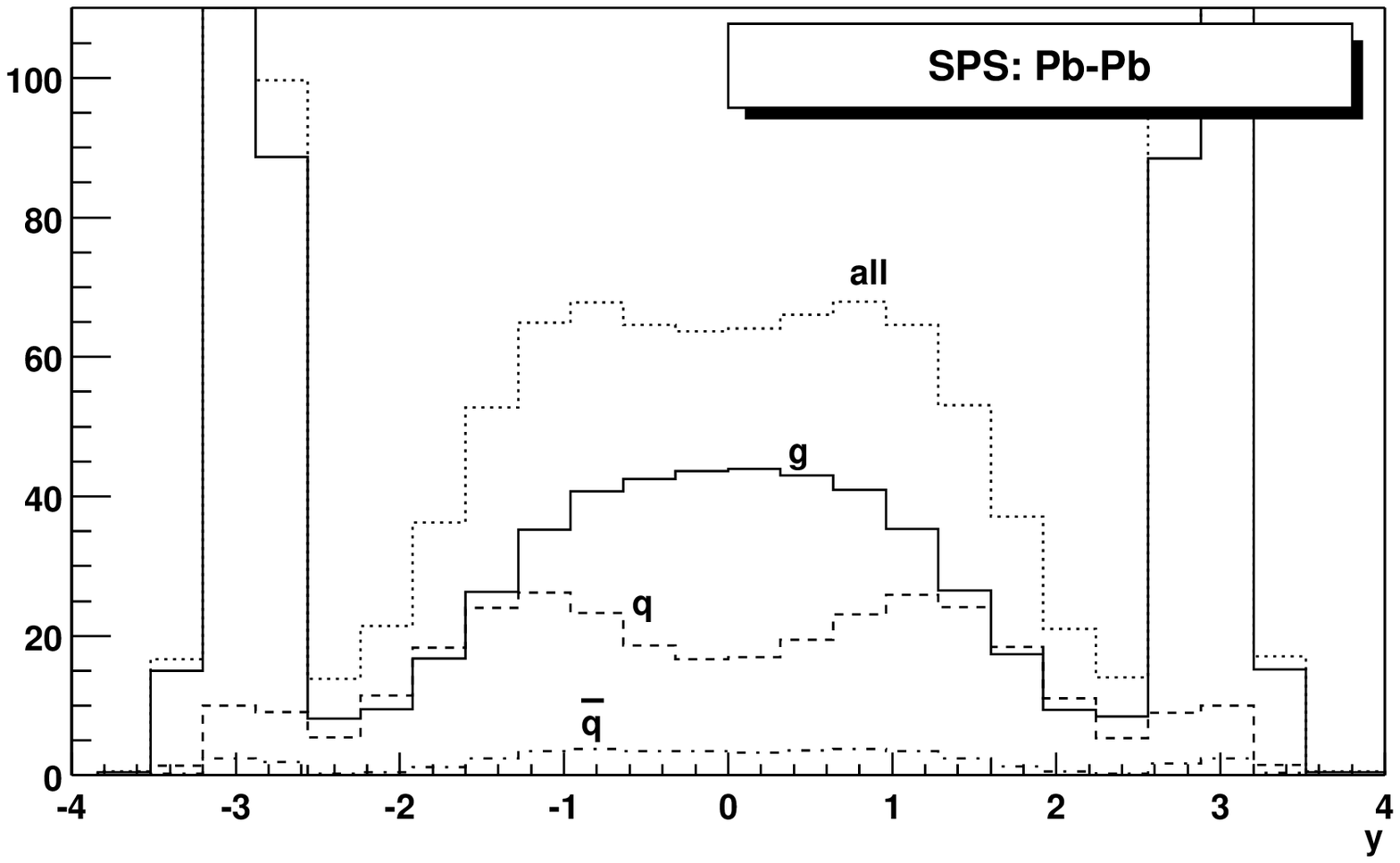}
                }
     \vspace{0.5ex}
     \caption{Final parton rapidity distributions for a S$-$S reaction
              ($\sqrt{s} = 2 \times 9.7 $ A$\cdot$GeV, top frame) and
              Pb$-$Pb ($\sqrt{s} = 2 \times 8.6 $ A$\cdot$GeV, bottom
              frame). Shown are the separate contributions of gluons
              ($g$), up and down quarks ($q$) and up and down antiquarks
              ($\bar{q}$).
             }               \label{fig:SPS-SS-Pb-hrap}
\end{figure}

In Fig.\ref{fig:SPS-SS-Pb-hrap} we show the final parton rapidity
distributions for S$-$S and Pb$-$Pb reactions. The contributions of the
most important flavors (gluons, quarks, antiquarks) are given separately
($q \equiv u+d+s+c$, $\bar{q} \equiv \bar{u}+\bar{d}+\bar{s}+\bar{c}$).

Our simulation reproduces the typical plateau at mid rapidity nicely.
Whereas the quarks and antiquarks show a dip in this region, the gluon
distribution is flat at mid rapidity, for Pb$-$Pb it almost shows a
small peak. This is due to the fact that predominantly gluons are
produced in binary parton$-$parton scatterings. Note that the peaks at
beam rapidity are mainly gluons; as in $p$$-$$\bar{p}$ they are probably
DGLAP gluons.

Fig.\ref{fig:SPS-net-baryon-hrap} shows the rapidity distribution of the
quantity $\sfrac{1}{3}(N_q -N_{\bar{q}})$ for the two reactions. In a
naive coalescence model along the lines of \cite{Bia98} or the ALCOR
model \cite{Bir95,Zim99} this quantity would be proportional to the net
baryon number. We therefore compare the above rapidity distributions
with the data of \cite{App99} (for the same reason as given above in the
case of $p$$-$$\bar{p}$, the scale is in arbitrary units). The agreement
is remarkable for both reactions, and (considering the larger errors
both in experiment and our calculation for S$-$S) seems better for the
larger system (Pb$-$Pb).

\begin{figure}
     \centerline{\includegraphics[keepaspectratio,width=75mm]%
                 {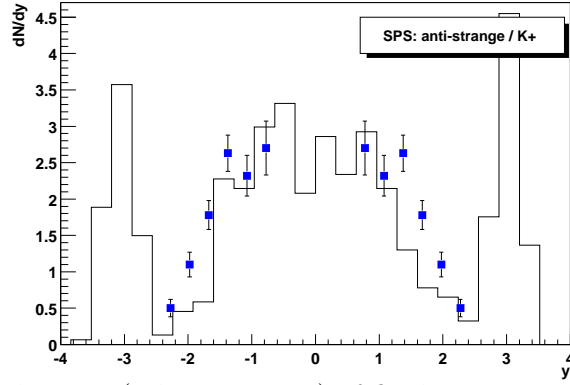}
                }
     \caption{Rapidity distributions (arbitrary units) of final
              antistrange quarks for a S$-$S reaction at $\sqrt{s} = 2
              \times 9.7 $ A$\cdot$GeV. The experimental data points are
              from \protect\cite{Alb98}.
             }               \label{fig:SPS-S-and-Pb-antistrange-hrap}
\end{figure}

Fig.\ref{fig:SPS-S-and-Pb-antistrange-hrap} presents the final
rapidity distributions of antistrange quarks for the S$-$S reaction
($\sqrt{s} = 2 \times 9.7 $ A$\cdot$GeV). Since the antistrange quarks
hadronize predominantly to $K^+$, we can compare their rapidity
distribution directly to the experimental anti-kaon rapidity
distribution, as given in \cite{Alb98} (no data are available for
Pb$-$Pb). The agreement of our simulation and the data is quite good
(apart from the peaks at beam rapidity, for which we have as yet no
convincing explanation).

\begin{figure}
     \centerline{\includegraphics[keepaspectratio,width=75mm]%
                 {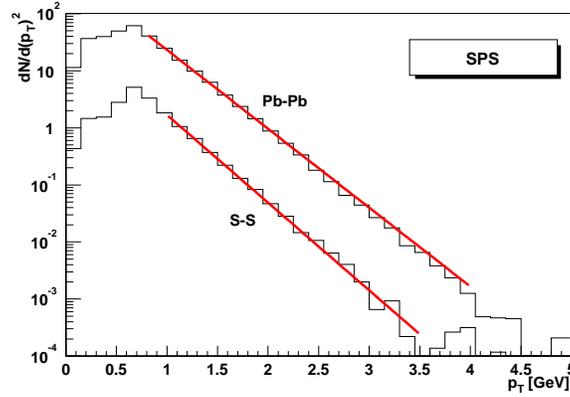}
                }
     \caption{Distributions of final parton transverse momenta for S$-$S
              ($\sqrt{s} = 2 \times 9.7 $ A$\cdot$GeV) and (Pb$-$Pb
              ($\sqrt{s} = 2 \times 8.6 $ A$\cdot$GeV). The solid lines
              merely serve to show to what extent the distributions are
              exponential.
            }                \label{fig:SPS-dndpt2}
\end{figure}

In Fig.\ref{fig:SPS-dndpt2} finally, we show the transverse momentum
distribution for both S$-$S and Pb$-$Pb reactions with the same energies
as before. The spectra show the same dip at low transverse momentum we
had in the $p-\bar{p}$ simulations (cf.\
Fig.\ref{fig:ppbar-pt-fit-200-1800GeV}).

Summarizing our comparison to the SPS S$-$S and Pb$-$Pb data, we find
that our model reproduces the rapidity and $p_{\mathsf{T}}$ spectra
surprisingly well.

\subsection{Au$-$Au at RHIC} \label{sec:RHIC}
We now present the results of PCPC simulations for RHIC physics: Au$-$Au
reactions at $2 \times 100$ A$\cdot$GeV.
We simulated 200 events. On the
average, 9300 partons per event were generated, of which 72\% were
`participating partons'. Again, only the latter are included in our
distributions.

\begin{figure}
     \centerline{\includegraphics[keepaspectratio,width=75mm]%
                 {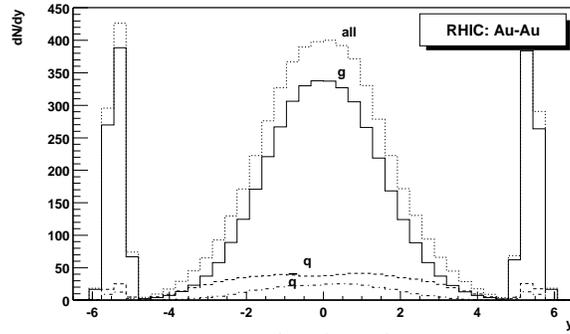}
                }
     \caption{Final parton rapidity distributions for Au$-$Au reactions
              at RHIC. Shown are the total partons and the separate
              contributions of gluons ($g$), quarks and antiquarks.
             }               \label{fig:RHIC-Au-rap}
\end{figure}

As before, we first present the results for final rapidity
distributions: Fig.\ref{fig:RHIC-Au-rap} contains the parton rapidity
distributions for all participating quarks, for gluons and for quarks
and antiquarks. As expected, the distributions are much more sharply
peaked than Fig.\ref{fig:SPS-SS-Pb-hrap}. Note in particular that the gluon
distribution (disregarding the peaks for the initial rapidities) is of
almost perfect Gaussian shape, and, in contrast to the SPS case, the dip
in the quark and antiquark distributions at mid rapidity has all but
disappeared. The ratio of gluons to quarks (at mid rapidity) is about
7:1, so that the mid rapidity region is a region of high energy density
and essentially baryon-free.

\begin{figure}
     \centerline{\includegraphics[keepaspectratio,width=75mm]%
                 {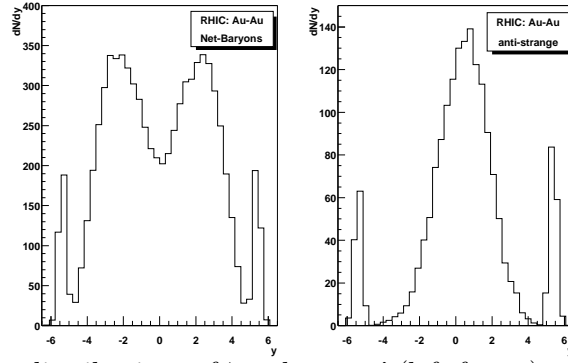}
                }
     \caption{Final rapidity distributions of `net baryons' (left
              frame) and antistrange quarks (right frame) for Au$-$Au
              reactions at RHIC.
             }          \label{fig:RHIC-net-baryon-and-antistrange-hrap}
\end{figure}

Although here we have no experimental data to compare to, we are again
interested in the quantity $\sfrac{1}{3}(N_q -N_{\bar{q}})$ as a measure
of the `net baryons' and the antistrange quarks as a measure of the
produced $K^+$. These are given in
Fig.\ref{fig:RHIC-net-baryon-and-antistrange-hrap} (regarding the beam
rapidity peaks, cf.\ the corresponding remarks in Sect.\ \ref{sec:SPS}).

\begin{figure}
     \centerline{\includegraphics[keepaspectratio,width=75mm]%
                 {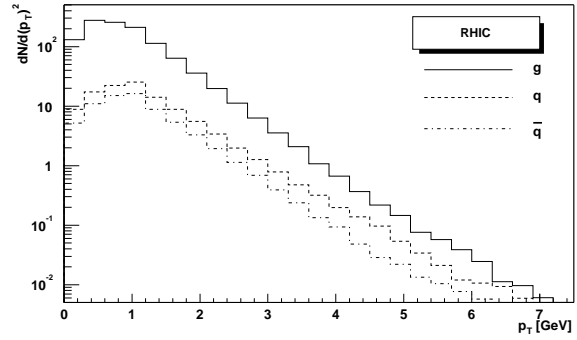}
                }
     \caption{Distributions of final parton transverse momenta
              for Au$-$Au reactions at RHIC.
            }                \label{fig:RHIC-dndpt2}
\end{figure}

Finally, we present in Fig.\ref{fig:RHIC-dndpt2} the distribution of
final parton transverse momenta. It is seen that the contributions of
all flavors of quarks and antiquarks are at least an order of magnitude
smaller than those of the gluons. Note that the dip at the lowest
$p_{\mathsf{T}}$ (again due to the neglect of soft interactions) is markedly
less for the RHIC Au$-$Au reactions than for $p$$-$$\bar{p}$ (cf.\
Fig.\ref{fig:ppbar-pt-fit-200-1800GeV}). Apart from this dip, the
distributions are nearly exponential.

All these results verify the expectation which we have theoretically:
that at RHIC energies we expect hard parton scatterings to play a much
more important r\^ole, and that therefore our model should be best
suited for that energy regime.
\section {Discussion and Conclusions}\label{sec:conclusions}

In this paper, we have presented a new parton cascade model which
differs from other such models in that it treats not only the kinematics
of the reaction, but all of the dynamics in a strictly
Poincar{\'e}-covariant manner.

In the light of the success of various other parton cascade models
(notably the VNI code, \cite{Gei92,Gei95}) this may seem to be a
merely formal aspect. There are, however, several practical advantages
in our covariant formulation:
\begin{itemize}
\item the algorithm (and the \emph{sequence of binary parton
      interactions}, in particular) does not depend on the frame of
      reference in which the code is run,
\item our model allows for a very natural treatment of parton off-shell
      effects (``virtualities'') which are included ad hoc in other
      models,
\item there is no need to use seemingly artificial mechanisms such as a
      ``distributed Lorentz contraction'' (cf.\ Sect.\ \ref{sec:dlc}) in
      order to enlarge the longitudinal extension of a nucleus before
      the collision. In fact, such a mechanism is inconsistent with our
      approach of insisting on strict Poincar{\'e}-covariance.
\end{itemize}

On the formal side, a fundamental problem with any cascade approach
remains also in PCPC. In Sect.\ \ref{sec:dynamic}, we have
explained how, in circumventing the No-Interaction-Theorem, we are led
to employ a many-times formalism. The invariant dynamical evolution
parameter $s$ of PCD, though consistently defined, does not lend
itself easily to a physical interpretation. As a consequence, the naive
idea of defining particle, energy and entropy densities in terms of
averages at given values of $s$ is not feasible, and indeed the problem
of formulating consistently the Poincar{\'e}-covariant statistical
mechanics of a system of \emph{classical} particles remains unsolved
(cf.\ \cite{Sor87a}).

It thus seems that we are defeating the very incentive for theoretically
modelling a reaction with a cascade code, viz.\ `looking inside the
reaction' \emph{during} the `hot and dense stages' of the collision.
This, however, is \emph{not} true. Quite to the contrary: \emph{because}
it is a Poincar{\'e}-covariant model, PCPC -- in contrast to
non-covariant models -- allows us to use the full phase space
information consistently to reconstruct the microscopic state of the
system as `seen' from any given observer frame at any given physical
(observer) time. Such a reconstruction, of course, provides only a formal
picture of the \emph{model}, not of physical reality: it must be pointed
out that not only is this information inaccessible to direct observation
in experiment, but the idea of actually \emph{looking simultaneously} at
the whole of a spatially extended system at \emph{one point in
(observer) time} is necessarily inconsistent with relativity,
irrespective of the particular formalism used by the theorist.

This -- theoretical -- visualization of the intermediate stages of the
reaction has in fact been quite useful to us in gaining insight, e.g.\
into the influence of the various parameters of our model (cf.\ Sects.\
\ref{sec:ini_state},\ref{sec:parton_scattering},\ref{sec:evolution}).
Wary of misinterpretation, however, we have refrained from presenting
such visualizations in this paper. Rather, we have restricted the
presentation of numerical results in Sect.\ \ref{sec:results} to
\emph{final distributions} which can be compared to experimental data
where such data are available (and the comparison is physically
meaningful).

\begin{figure}
     \centerline{\includegraphics[keepaspectratio,width=75mm]%
                 {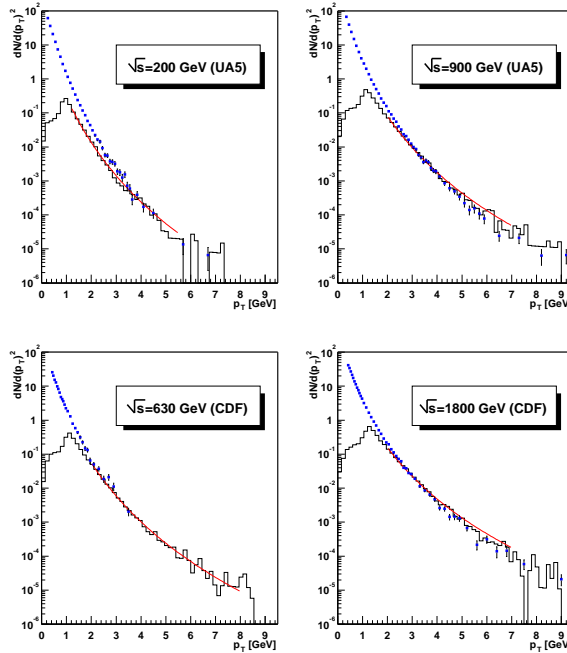}
                }
     \caption{Distributions of final transverse momenta for a
              $p$$-$$\bar{p}$ reaction at
              \protect\\
              $\sqrt{s}= 200,630,900 \text{ and } 1800 \text{ GeV}$.
              Note the rapidity cut: only particles with $|y|<1$ are
              included. The histograms are the PCPC results. The data
              points (including the fit lines through them) are the
              experimental $\frac{dN_{ch}}{dp_{\,\mathsf{T}}}$ (from
              \protect\cite{Alb90} and \protect\cite{Abe88}).
             }               \label{fig:ppbar-pt-fit-200-1800GeV}
\end{figure}

\begin{figure}
     \centerline{\includegraphics[keepaspectratio,width=75mm]%
                 {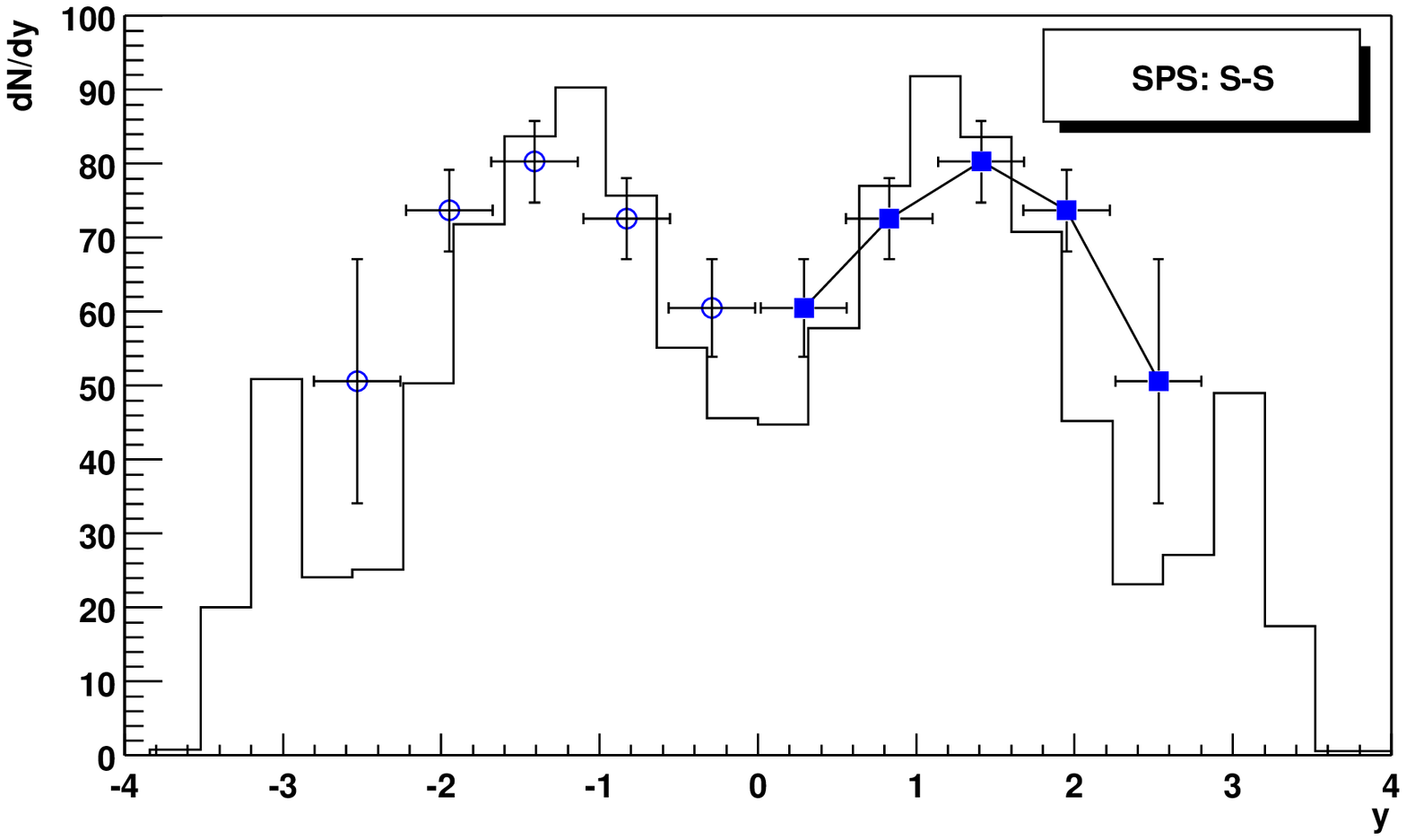}
                }
     \vspace{6pt}
     \centerline{\includegraphics[keepaspectratio,width=75mm]%
                 {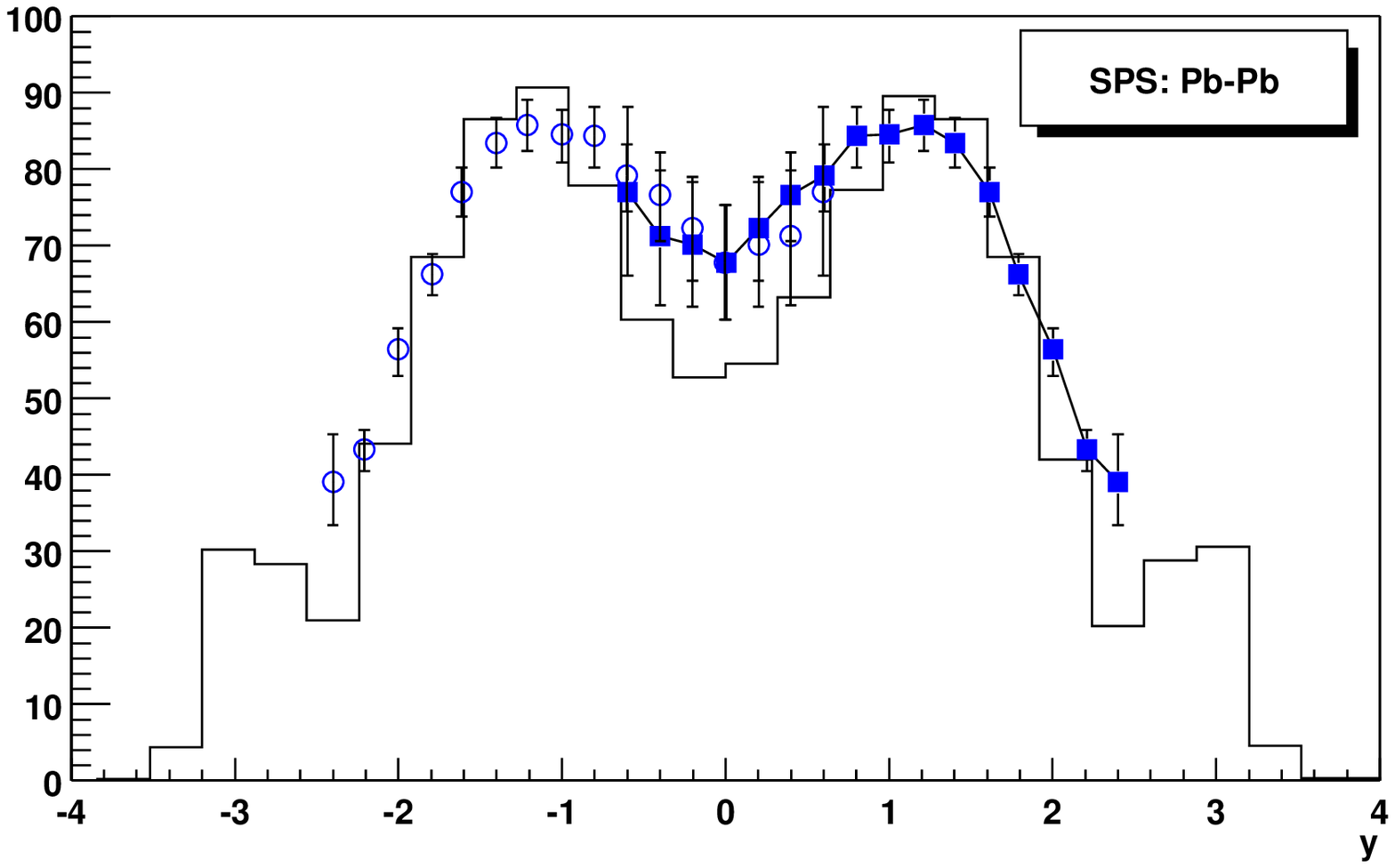}
                }
     \caption{Final rapidity distributions of `net baryons' (arbitrary units) for a S$-$S
              reaction ($\sqrt{s} = 2 \times 9.7 $ A$\cdot$GeV, top
              frame) and Pb$-$Pb ($\sqrt{s} = 2 \times 8.6 $
              A$\cdot$GeV, bottom frame). The histograms are results of
              the PCPC simulation, the data point and their error bars
              are from \protect\cite{App99}.
             }               \label{fig:SPS-net-baryon-hrap}
\end{figure}

In our view, these comparisons show that PCPC simulates the reactions
reasonably well. In particular, we want to draw attention again to
Figs.\ \ref{fig:ppbar-pt-fit-200-1800GeV} and
\ref{fig:SPS-net-baryon-hrap}: Fig.\ \ref{fig:ppbar-pt-fit-200-1800GeV}
shows that the agreement improves for higher energies, and in Fig.\
\ref{fig:SPS-net-baryon-hrap} we see that PCPC does better for heavier
systems. This is precisecly what one would expect, and it strengthens
our belief that PCPC will be useful in the RHIC regime.

Another way to assess the usefulness of PCPC is to compare its results
with comparable theoretical models. Such a comparison of our results
with those of VNI has been presented elsewhere \cite{Boe99}.

In the present version, PCPC contains no hadronization scheme. As was
pointed out before (cf.\ Sect.\ \ref{sec:results}), we feel that in a
heavy ion reaction a hadronization mechanism which is added in the final
stage (i.e.\ after the parton cascade has come to its end) is somewhat
artificial, and phenomenological at best. What we envisage is an
\emph{integrated hadron-parton cascade}, in which partons are formed in
binary \text{scatterings} of the initial nucleons, and hadrons are
formed (and `dissolved' again) continually while the reaction is going
on. In such a model, the initial state of the system would be
constructed quite naturally of nucleons only, thus removing some of the
artificialities described in Sect.\ \ref{sec:ini_state}. This, however,
is work that remains to be done.

The PCPC code [in $\text{C}^{++}$] is obtainable from the OSCAR archive
\begin{center}
\texttt{http://rhic/phys.columbia.edu/oscar})
\end{center}
 or from the authors.

\acknowledgements

The research presented here would not have been possible without the
seminal input provided by its non-covariant forerunner, the VNI code
by Klaus Geiger. We deeply regret that our work can no longer be
subjected to his productive criticism. We dedicate this paper to the
memory of Klaus Geiger.



    \renewcommand{\arraystretch}{1.2}
\begin{table}                   
  \begin{center}
    \begin{tabular}{l|cc}
        process & $|{\mathcal{M}}/g^2|^2$               \\ \hline\hline
        $gg\rightarrow gg$
                & $G_0(s,t,u)$                           \\ \hline
        $gg\rightarrow q\bar{q}$
                & $G_3(s,t,u;m)$                         \\ \hline
          $\!\!\begin{array}{l}
                   qg \rightarrow qg ,        \\
             \bar{q}g \rightarrow \bar{q}g
           \end{array}$
                & $\left\langle\sfrac{8}{3}\right\rangle G_3(t,s,u;m)$
                                                         \\ \hline
          $\!\!\begin{array}{l}
                         qq \rightarrow qq , \\
             \bar{q}\bar{q} \rightarrow \bar{q}\bar{q}
           \end{array}$
                &
          $\!\!\begin{array}{r} G_1(s,t,u;m,m) + G_1(s,u,t;m,m) \\
                           +G_2(s,t,u;m)
           \end{array}$                                \\ \hline
        $q\bar{q} \rightarrow q\bar{q}$  \rule[-2ex]{0pt}{4ex}
                &
          $\!\!\begin{array}{r} G_1(u,s,t;m,m) + G_1(u,t,s;m,m) \\
                           +G_2(u,s,t;m)
           \end{array}$                                \\ \hline
        $q\bar{q} \rightarrow q^\prime{\bar{q}}^\prime$
                & $ G_1(u,s,t;m,m^\prime)$             \\ \hline

          $q\bar{q} \rightarrow gg$       \rule[-2ex]{0pt}{5ex}
                & $\left\langle\sfrac{64}{9}\right\rangle G_3(s,t,u;m)$
                                                        \\ \hline
          $\!\!\begin{array}{l}
                          qq^\prime \rightarrow qq^\prime , \\
                   q{\bar{q}}^\prime \rightarrow q{\bar{q}}^\prime , \\
             \bar{q}{\bar{q}}^\prime \rightarrow \bar{q}{\bar{q}}^\prime
           \end{array}$
                & $G_1( s,t,u;m,m^\prime)$
    \end{tabular}
    \caption{Matrix elements (with the coupling constant factored out)
             for the partonic processes $ab\rightarrow cd$. The
             numerical factors in brackets are due to different colour
             averages.
             \label{tab:subprocesses}
            }
  \end{center}
\end{table}

\begin{table}                      
  \renewcommand{\arraystretch}{1.5}
  \begin{center}
    \begin{tabular}{c}
      $ \begin{array}{l}
          \frac{4}{3} \frac{1+z^2}{1-z}    \\
          \frac{4}{3} \frac{1+(1-z)^2}{z}  \\
          \text{\footnotesize{6}} \,
          \frac{[1-z(1-z)]^2}{z(1-z)}      \\
          \sfrac{1}{2}  \left[ z^2 + (1-z)^2      \right]
        \end{array}
        \text{\qquad for }
        \left\{ \begin{array}{l}
                   q\rightarrow qg         \\
                   q\rightarrow gq         \\
                   g\rightarrow gg         \\
                   g\rightarrow q\bar{q}
                \end{array}
        \right.
      $
    \end{tabular}
  \end{center}
  \caption{Altarelli-Parisi splitting functions
           $P_{a\rightarrow bc}(z)$ . From \protect\cite{Alt77}.
          }                             \label{tab:APsplitting}
\end{table}

\begin{table}                      
  \renewcommand{\arraystretch}{1.2}
  \begin{center}
    \begin{tabular}{c|cc}
      Flavor $b$  & Flavor $a$                           \\ \hline
          $g$     &     $g$                              \\
                  &$u,\bar{u},d,\bar{d},s,\bar{s}\ldots$ \\ \hline
     $u(\bar{u})$ &$u(\bar{u})$                          \\
                  &     $g$                              \\ \hline
     $d(\bar{d})$ &$d(\bar{d})$                          \\
                  &     $g$                              \\ \hline
     $s(\bar{s})$ &$s(\bar{s})$                          \\
                  &     $g$                              \\ \hline
       $\vdots$   &  $\vdots$
    \end{tabular}
    \caption{Flavor generation in the parton evolution. A parton with
             flavor $b$ may be radiated from a parton with flavor $a$.
             The list includes all possible QCD branchings.}
    \label{tab:processes}
  \end{center}
\end{table}

\begin{table}                      
  \renewcommand{\arraystretch}{1.5}
  \begin{center}
    \begin{tabular}{lc|cccc|c|c|c}
                                && \multicolumn{4}{c|}{$p$$-$$\bar{p}$}
                                    & S$-$S & Pb$-$Pb & Au$-$Au     \\
           $\sqrt{s}$           & [A$\cdot$GeV]
           & 200 & 615 & 900 & 1800
                             & 9.7
                             & 8.6
                             & 100 \\ \hline
           $x_{\text{min}}$     &
           & 0.01 & 0.003 & 0.002 & 0.001 & 0.1 & 0.12 & \\
           ${Q_0}^2$            & [$\text{GeV}^2$]
           & 5.3 & 9.8 & 11.7 & 16.6 & 1.6 & 1.3 & 5.3 \\
           ${Q_{\text{min}}}^2$ & [$\text{GeV}^2$]
           & 1.73 & 2.36 & 2.60 & 3.14 & 0.92 &0.89 & 1.73 \\
    \end{tabular}
  \end{center}
  \caption{Values of various parameters used in the numerical
           computations (the meaning of these parameters is explained in
           Sects.\ \ref{sec:ini_state} and \ref{sec:parton_scattering}).
          }                             \label{tab:parameters}
\end{table}

\end{document}